\documentstyle[12pt]{article}
\newlength{\extraspace}
\newlength{\extraspaces}
\newcommand{\bq}{\begin{eqnarray}
\addtolength{\abovedisplayskip}{\extraspaces}
\addtolength{\belowdisplayskip}{\extraspaces}
\addtolength{\abovedisplayshortskip}{\extraspace}
\addtolength{\belowdisplayshortskip}{\extraspace}}
\newcommand{\eq}{\end{eqnarray}}
\begin{document}
\addtolength{\baselineskip}{.8mm}
\thispagestyle{empty}

\begin{flushright}
{\sc OUTP}-96-43 P\\
{\sc TPI-MINN}-96/11-T\\
\today \\
 hep-th/9608031
\end{flushright}
\vspace{.3cm}

\begin{center}
{\large\sc{ THREE-DIMENSIONAL DESCRIPTION OF THE 
 $\Phi_{1,3}$ DEFORMATION OF MINIMAL MODELS} }\\[15mm].

{\sc  Ian I. Kogan}\footnote{e-mail address:
i.kogan1@physics.ox.ac.uk }\footnote{ On  leave of absence
from ITEP,
 B.Cheremyshkinskaya 25,  Moscow, 117259, Russia}
\\[2mm]
{\it Theoretical Physics, University of Oxford, 1 Keble Road,
       Oxford, OX1 3NP, UK }\footnote{permanent address}\\[5mm]
 {\it and} \\[5mm]
{\it Theoretical Physics Institute, 
 Physics Department, University of Minnesota\\
116 Church st. S.E., Minneapolis, MN 55455, USA }\\[15mm]

{\sc Abstract}
\end{center}

\noindent
 We discuss the $2+1$ dimensional description of  the $\Phi_{1,3}$
 deformation of the minimal model $M_p$ leading to 
 a  transition $M_p \rightarrow M_{p-1}$. The deformation  can be
 considered as an addition of the charged matter to the Chern-Simons theory
 describing a  minimal  model. The $N=1$  superconformal case is also
 considered.
\vfill
\newpage
\pagestyle{plain}
\setcounter{page}{1}

It is  known that two-dimensional conformal field theories 
 ($2D$  CFT)  \cite{CFT},  \cite{book}
can be described in three-dimensional terms  by  using an amusing connection
 \cite{Wita}  between  a $2+1$-dimensional 
 topological  Chern-Simons (CS) theory with an action 
\begin{equation}
kS_{CS}\{A\} = {k\over4\pi}\int_{\cal M} Tr \left(A\wedge dA +
{2\over3}A\wedge A\wedge A\right)
\label{csaction}
\end{equation}
defined on a three-dimensional
 manifold $M$ with a boundary $\Sigma = \partial M$  and 
 a  Wess-Zumino-Novikov-Witten (WZNW) model on a two-dimensional
  boundary $\Sigma$. 
     It is also known that  two-dimensional 
 conformal field theories can be considered as fixed points 
(infrared or ultraviolet) of the  renormalization group  (RG) flows in 
the space of all
 $2D$ renormalizable quantum field theories (QFT). In this approach,
 which was  initiated in \cite{zam1}, \cite{cl}, a conformal field
theory with an action $S_{CFT}$  is deformed by  some operators $V_i$
 with  scaling dimensions $d_{i}= \Delta_{i} + \bar{\Delta}_{i}$
 and the action is
\begin{equation}
S[\mu] = S_{CFT} + \sum_{i} \lambda^i[\mu] \int d^2\xi ~V_{i}(\xi)
\label{deformedaction}
\end{equation}
 where coupling constants $\lambda^i$  depend on a scale $\mu$
 as well  as the whole action (\ref{deformedaction}). In the case of a
 relevant  deformation $d_{i} < 2$ the coupling constants vanish in the 
 ultraviolet (UV) limit  and one can recover conformal field theory
 $S_{CFT}$ as an UV fixed point. In the case of an irrelevant deformation
 $d_{i} >2$
 the  coupling constants vanish in the infrared (IR) limit and one gets 
 conformal field theory $S_{CFT}$ as
 an infrared fixed point of the renormalization group flow.
 
  If the renormalization group exibits topologically nontrivial behaviour,
 i.e. has another fixed point in the vicinity of the original one, there
 is a  RG flow connecting {\it two} different conformal field theories.
 The toy model of this phenomenon is the famous RG flow
 $M_k \rightarrow M_{k-1},~k = 1,2,3,4,\dots$  \cite{zam1} \cite{cl} where the 
 action (\ref{deformedaction}) interpolates between the UV fixed point
 $M_{k}$ and  the IR fixed point $M_{k-1}$ and the deformation operator
 $V$ is a $\Phi_{1,3}$ field  with anomalous dimensions
  $\Delta_{1,3} = \bar{\Delta}_{1,3} = 1 - 2/(k+3)$  in  a vicinity
 of the UV  fixed point $M_{k}$. Let us note that we are using here nonstandard
 notation, the standard one  for minimal models is 
  $M_{p},~~p = k+2 = 3,4,...$.

 Minimal models  can be described by the 
 $SU(2)_{k} \times SU(2)_{1}/ SU(2)_{k+1}$ GKO coset construction
\cite{gko} and  a corresponding
 Chern-Simons  description is given by three $SU(2)$ gauge fields 
 $A$, $B$  and $C$ with action \cite{ms}
\begin{equation}
kS_{CS}\{A\} + S_{CS}\{B\} - (k+1)S_{CS}\{C\}
\label{cscoset}
\end{equation} 
 We shall use obvious notation $[k,1 ; k+1] = [k, 1, -(k+1)]$ for the triplets
 of CS coefficients.
 For the $M_k \rightarrow M_{k-1}$ flow  one has   {\it two different}
 Chern-Simons theories corresponding to the UV and IR fixed points of
 the flow. One can ask immediately
  the following questions:
\begin{itemize}
\item {\sl  What is necessary to add to a  $2+1$-dimensional topological
CS theory to be able to get a three-dimensional  description  of a 
  deformed conformal field theory ?}
\item  {\sl How will this new ingredient change the original CS theory
 corresponding
 to the $M_{k}$ model into a new one which must describe the $M_{k-1}$ model 
in the
 infrared limit?} 
\end{itemize}

The answer to the first question (in general) is known \cite{kogan} - 
 we have to add charged matter, so that the three-dimensional theory 
  is not 
 topological anymore - there are propagating degrees of freedom in the bulk. 
   But it was  unclear  how to add  a charged matter in such a way that
 in the infrared limit the new CS theory will arise. The aim of this letter
 is to answer the second question and to discuss the  three-dimensional
 description of the two-dimensional RG flow $M_{k}\rightarrow M_{k-1}$ 
 as well as it SUSY  generalization.

Let us remind  how a charged matter added in the bulk  will induce 
  a deformed $2D$ CFT on the boundary (for details see \cite{kogan}).
 The statistical sum of 
  a  deformed conformal field theory ( in a critical string theory
  it gives  a generating function in an  external field)  is
 determined by the action (\ref{deformedaction}) and in case of a  single-charge
 deformation  is given by 
\begin{eqnarray}
 Z = \int D \Phi(\xi) exp [ -S_{CFT} + \lambda \int d^2\xi ~V(\xi, \bar{\xi})]
= \nonumber \\
\sum_{n} \frac{\lambda^{n}}{n!}\int
d^{2}\xi_{1}\cdots\int d^{2}\xi_{n}\langle V(\xi_{1},\bar{\xi}_{1})\cdots
V(\xi_{n}, \bar{\xi}_{n})\rangle, 
 \label{ssum}
 \end{eqnarray}
where the brackets $\langle V(\xi_{1},\bar{\xi}_{1})\cdots
V(\xi_{n}, \bar{\xi}_{n})\rangle$ are $n$-point  correlation functions in  an 
 unperturbed CFT which can be represented as products of left
 and right conformal blocks $\langle V(\xi_{1},\bar{\xi}_{1})\cdots
V(\xi_{n}, \bar{\xi}_{n})\rangle =
 \langle V_{L}(\xi_{1})\cdots
V_{L}(\xi_{n})\rangle~\langle V_{R}(\bar{\xi}_{1})\cdots
V_{R}(\bar{\xi}_{n})\rangle$ where $V_{L}(\xi)$ and $V_{R}(\bar{\xi})$ are 
holomorphic and antiholomorphic 
  chiral vertex operators corresponding to the left-right symmetric operator 
$V(\xi, \bar{\xi})$. To obtain a three-dimensional  picture we have to consider
 a   membrane with topology  $M = \Sigma \times I$ where the two
 boundaries $\Sigma_{L}$ and $\Sigma_{R}$ are connected by an interval 
$I$ of a 
 length $\beta$. A gauge theory with a  Chern-Simons term in the bulk  will
 induce left and right sectors of a $2D$ CFT (actually it depends on boundary
 conditions for gauge fields, 
for more details see \cite{leith}) and an insertion of the
  vertex operator
 $V(\xi, \bar{\xi})$  on a world-sheet $\Sigma$ is equivalent to  insertions
 of chiral vertex operators $V_{L}(\xi)$ and $V_{R}(\bar{\xi})$ on  left
 and right world-sheets $\Sigma_{L}$ and $\Sigma_{R}$ respectively  with the
 coordinates on both world-sheets identified in the end. This
  is induced by an open
 Wilson line 
\begin{equation}
W_{\{R_{k}\}}(C_{\xi_{L}, \xi_{R}})
 =\prod_{k} ~Tr_{R_{k}}~\exp\left(i\int A^{(k)}_{\mu}dx^{\mu}\right)
\label{wilsonline}
\end{equation}
along the path $C_{\xi_{L}, \xi_{R}}$ with end points $\xi_{L}$ and $ \xi_{R}$
 on left $\Sigma_{L}$ and right $\Sigma_{R}$ world-sheets respectively.
 \cite{Wita}, \cite{ms}-\cite{membrane}. The insertion of this Wilson line 
in the bulk
 gives  the phase factor of a propagating  charged
  particle where  charges with respect to gauge fields $A^{(k)}$ are
 given by representations $R_{k}$ (this set of quantum numbers  depends on  the
 type of vertex operator under consideration). The quantum particle
  propagates  from 
 left to right world-sheets and 
  a gas of these 
open Wilson lines describes  a charged matter in the bulk. The  third dimension
 along the interval $I$ 
 plays the role of an imaginary time  and the  
 parameter $\beta$ (internal size of a  membrane)   can be interpreted
 as an inverse temperature $\beta = T^{-1}$. In this way   the connection 
 between  a  charged $2+1$-dimensional matter at a temperature $T$ and a 
deformed
  two-dimensional conformal field theory is established.
  
It is easy to see that the fugacity $\lambda$  depends on the temperarture 
$T = 1/\beta$.
 The Wilson line (\ref{wilsonline}) is a phase factor in
  the path integral describing the propagation of the quantum particle with
 mass
 $m$  (let us for simplicity consider the simplest case of scalar particle,
 in other
 cases the conclusion will be the same)  from $\Sigma_{L}$ to $\Sigma_{R}$
\begin{eqnarray}
G(\xi_{L}, \xi_{R}) = \int_{0}^{\infty} d\tau 
\exp\left[-\frac{1}{2}m^{2}\tau\right]
\int_{x(0) = \xi_{L}}^{x(\tau) =  \xi_{R}}
{\cal D}x(t)
\exp\left[-\frac{1}{2}\int_{0}^{\tau}dt~\dot{x}^{\mu}\dot{x}_{\mu}\right]
W_{\{R_{k}\}}(C_{\xi_{L}, \xi_{R}})
\end{eqnarray}
 where $x^{\mu}(t)$ is a three-dimensional  
coordinate along a quantum path  and $\tau$ is the proper time. The classical
 path
 is a straight line $x^{\mu}(t) = \delta^{\mu,3} \beta ~t/\tau$ 
 and an extremal
 value of the proper time is $\tau = \beta/m$ from which one gets the leading
 (classical) contribution $\exp[-\beta ~m]W_{\{R_{k}\}}(C_{\xi_{L}, \xi_{R}})$
 and each Wilson line is accompanied by a fugacity factor
 $\lambda \sim \exp[-\beta ~m]$. In the low-temperature limit $(T/m) 
\rightarrow 0,~
 (m \beta) \rightarrow \infty$ the fugacity, i.e. deformation parameter
 $\lambda$
  disappears and  we have a conformal field theory. It is necessary to have
 in mind
 that the same charge matter can, in principle, renormalize the parameters
 of the
 Chern -Simons terms. The problem is that besides the Wilson lines connecting
 left and right world-sheets (using our analogy with finite temperature
one can see that they are nothing
 but Polyakov's lines) there are ordinary closed loops in the bulk
 which do not touch  the boundary. These last ones do not induce any vertex
 operators
 insertions  and are not supressed in the low $T$ limit. They describe  
production
  and annihilation of virtual  pairs and any vacuum loops will renormalize
 parameters of the gauge theory. If the charged matter is $P$-odd, then these
  loops   give contributions to the total Chern-Simons coefficients which 
become
 $T$- dependent. Thus we see that the bulk parameters will also experience some
 kind of flow from small $T$ to large $T$.  At the same time the fugacity
 $\lambda(T)$ starts to increase and as a result the coupling constant of the
 induced two-dimensional theory  starts to increase. In the limit of 
infinite $T$
  or zero $\beta$ we can reach (if we have matter with suitable charges) a new
 set of Chern-Simons coefficients describing another conformal field theory.
 In such a way we obtain the  three-dimensional (membrane or bulk) description
  of a two-dimensional RG flow. Let us note that  a matter contribution to
  the initial set of
 the  CS coefficients  is calculated at zero (or very low) temperature 
 $T << m$. The
 same contribution to the final set (corresponding to the IR fixed point)
 must be calculated  in the high-temperature limit $T >> m$, in which case
 the matter contribution will be proportional to $\tanh (m/T) \rightarrow 0$
 and can be neglected.

 Let us apply now these ideas to 
  $M_{k}\rightarrow M_{k-1}$ flow.  It seems that we have to  demonstrate
 that matter contribution must change Chern-Simons coefficients in 
(\ref{cscoset}) from $(k-1,1;k)$ into $(k,1;k+1)$. To see if this is
 possible the quantum numbers of the $\Phi_{1,3}$ operator must be identified
  first of all.
 As was demonstrated by Goddrad, Kent and Olive \cite{gko} 
  representations of the affine
 Kac-Moody algebra $\widehat{SU(2)}_{k} \times \widehat{SU(2)}_{1}$ can
  be decomposed
 with respect to $\widehat{SU(2)}_{k+1} \times V(c)$, where  $V(c)$  denotes
 the Virasoro algebra of the minimal model $M_{k}$  with central charge
\begin{equation}
c_{M_{k}} = c_{SU(2)_k}+c_{SU(2)_1}-c_{SU(2)_{k+1}} = 1 -\frac{6}{(k+2)(k+3)}
\label{minmodcentrcharge}
\end{equation}
Highest weight irreducible  unitary representations of $\widehat{SU(2)}_{k}$
  are  labelled by $(k, l)$ and called level $k$, spin $l$ representations,
 where $l$ is the spin ($SU(2)$ charge) of the
 corresponding primary field and the restriction  $0 \leq 2l \leq k$ 
must apply. 
 The  product of two representations $(k,l)$ and $(1,\epsilon)$ decomposes into
 the direct sum
\begin{equation}
(k,l) \times (1,\epsilon) = \bigoplus_{l'} \left(k+1, \frac{1}{2}(q-1)\right)
 \times (c, \Delta_{2l+1,2l'+1}(c))
\label{decomposition}
\end{equation}
where $c$ is given by (\ref{minmodcentrcharge})  and 
\begin{equation}
\Delta_{p,q}(c) = \frac{[(k+3)p -(k+2)q]^2 - 1}{4(k+2)(k+3)}
\label{Deltapq}
\end{equation}
  are anomalous dimensions of the primary fields with respect to the Virasoro
  algebra $V(c)$. 
The sum in  (\ref{decomposition}) is taken over $l'$  such that  $2(l-l')$ 
is even or odd, 
 depending on whether $\epsilon = 0$ or $1/2$, and $1 \leq q = 2l'+1 \leq k+2$.
 The decomposition (\ref{decomposition}) implies the following relations
 between
 characters
\begin{equation}
\chi_{k,l}(q,\theta)~\chi_{1,\epsilon}(q,\theta) =
 \sum_{l'} \chi_{k+1, l'}(q,\theta) \chi^{V}_{c,\Delta_{2l+1,2l'+1}} (q)
\label{charactersdecomposition}
\end{equation}
 where 
\begin{eqnarray}
\chi_{k,l}(q,\theta) = Tr \left(q^{L_{0}} exp(i\theta T^3) \right)
 = q^{l(l+1)/(k+2)} \times \nonumber \\
\frac{\sum_{n\in Z}
q^{n^2(k+2) + n(2l+1)}\left\{\exp(i(l+n(k+2))\theta) -
\exp(-i(l+1 + n(k+2))\theta) \right\} }
{\prod_{n=1}^{\infty}(1-q^n)(1-q^ne^{i\theta})(1-q^{n-1}e^{-i\theta})}
\label{kmcharacter}
\end{eqnarray}
and 
\begin{eqnarray}
\chi^{V}_{c,\Delta_{p,q}} (q) = Tr_{\Phi_{\Delta}}\left(q^{L_0}\right)
= \sum_{n\in Z}\left\{q^{\alpha(n)} - q^{\beta(n)}  \right\}
\prod_{n=1}^{\infty}(1-q^n)^{-1} \nonumber \\
\alpha(n) = \frac{[(k+3)p -(k+2)q +2n(k+2)(k+3)]^2 - 1}{4(k+2)(k+3)}; \\
\beta(n) = \frac{[(k+3)p +(k+2)q +2n(k+2)(k+3)]^2 - 1}{4(k+2)(k+3)}
\nonumber
\label{virasorocharacter}
\end{eqnarray}
are characters for Kac-Moody \cite{kac} and Virasoro \cite{rc} algebras respectively.

 Thus to get the field $\Phi_{1,3}$ we must take  $ l = 0$, $\epsilon = 0$
  and $l' = 1$. Let us also mention that the field $\Phi_{3,1}$ into which
   the first one must flow when approaching the IR fixed point $M_{k-1}$,
  must have $l =1$,  $\epsilon = 0$ and $l' = 0$. In spite of the fact that
 $\epsilon = 0$ it is wrong that the $\Phi_{1,3}$ field has no charge
 with respect to $SU(2)_{1}$. It is easy to see using a very simple expression
 for the Kac-Moody character at level $1$
 \begin{equation}
 \chi_{1,\epsilon}(q, \theta) = \sum_{m\in Z+\epsilon} q^{m^{2}} e^{im\theta}
 \prod_{n=1}^{\infty}(1-q^n)^{-1}
 \label{km1}
\end{equation}
  and 
  (\ref{charactersdecomposition}), (\ref{kmcharacter}) and  
(\ref{virasorocharacter}) that    for $l'=l+n$ ($n$ is an integer for
  $\epsilon  = 0$ and a half-integer for  $\epsilon  = 1/2$)
 the leading terms from $\chi_{1,\epsilon}$ which contribute to the $l'$ term
 in the sum (\ref{charactersdecomposition})  are $q^{n^2}e^{\pm i n\theta}$.
 This means that $\Phi_{3,1}$ is a descendant of a unity operator 
 with respect to $SU(2)_{1}$ with a spin $n^2$. This also can be seen from the
 following representation of the anomalous dimension (\ref{Deltapq})
\begin{eqnarray}
\Delta_{2l+1,2l'+1} = \frac{[(k+3)(2l+1) -(k+2)(2l'+1)]^2 - 1}{4(k+2)(k+3)} =
 \nonumber \\
n^2 + \frac{l(l+1)}{k+2} - \frac{l'(l'+1)}{k+3};~~~~n = l'-l
\label{Deltafinal}
\end{eqnarray} 
which makes absolutely clear what part of the
anomalous dimension (\ref{Deltafinal})
 came from each of the three $SU(2)$ sectors. This representation has a  
  straightforward three-dimensional interpretation. It is known (see \cite{acks}
  and references therein) that anomalous dimensions of $2D$ conformal fields
 can be obtained from the
 Aharonov-Bohm part of  $2+1$ scattering amplitudes for corresponding  dynamical 
 matter fields. From  the formula (\ref{Deltafinal}) we see that the total
anomalous dimension (proportional to the total amplitude) is the sum of three
 independent contributions from $SU(2)_{1}$,  $SU(2)_{k}$  and
 $SU(2)_{k+1}$. The last one is negative, because of the negative sign
 of the last term in (\ref{cscoset}).

 The obtained quantum numbers 
 for $\Phi_{1,3}$ operator seem very strange, because the corresponding matter
 field  must be  charged with respect to $B$ and $C$ fields in (\ref{cscoset})
 but not with respect to the  $A$ field. There is no way the first coefficient
 $k$ will be influenced by presence of this matter field and at first sight it seems
 impossible to get what we have aimed for. There is a loophole, however.
 Let  us  give first a 
heuristic argument as to what  we  would like the renormalizations to be.  The
 fact that there is a charge with respect to $SU(2)_{1}$ makes it possible
 to renormalize the coefficient $1$ also. This second "negative" result actually
  means that if we shall flip the sign for $SU(2)_{1}$, i.e. 
 transform $1$ into $1-2 =-1$ for $B$  field and $k$
 into $k-2$ for $C$ field (leaving coefficient $k$ in front of CS action for
 $A$ field intact) we shall transform 
 $[k,1 ; k+1] = [k, 1, -(k+1)]$ into $[k;1,  k-1] = [k, -1, -(k-1)]$ which means
 that we practically get what we need exept the overall sign.     
By making a parity transformation (changing the orientation of the 3-manifold $M$) we
 can change the overall sign of all the CS terms (\ref{csaction}). 
As a result we have the following transformation: 
\begin{eqnarray}
kS_{CS}\{A\} + S_{CS}\{B\} - (k+1)S_{CS}\{C\} \stackrel{\rm Renorm.}
{\longrightarrow}
kS_{CS}\{A\} - S_{CS}\{B\} - (k-1)S_{CS}\{C\} \nonumber \\
 \stackrel{\rm Parity}{\longrightarrow}
(k-1)S_{CS}\{\tilde{A}\} + S_{CS}\{\tilde{B}\} - kS_{CS}\{\tilde{C}\};
~~~~ (A~,B~,C) \rightarrow (\tilde{C},~\tilde{B},~\tilde{A})~~~~~~~~~~ 
\label{renormcscoset}
\end{eqnarray}
 Before demonstrating that precisely this renormalization takes place let us see
  that the suggested picture of the sign flip of the  $B$ field  which  leads 
 to an effective  exchange  of  roles
  between the  $A$ and $C$ fields is consistent with  known
  renormalizations of the fields $\Phi_{n,m}$ in a  perturbed theory
 $M_{k,k-1} = M_{k}\rightarrow M_{k-1}$ \cite{zam1}.  
  To first order the
 mixing between operators is determined by the fusion rule
\begin{equation}
\Phi_{n,m}\times \Phi_{1,3} = \left[\Phi_{n,m}\right]+
\left[\Phi_{n,m-2}\right]+\left[\Phi_{n,m+2}\right]
 \end{equation}
where the square brackets denote the contribution of the corresponding operator
  and  of its descendants. This rule is a consequence of the fact that $l=0$ and
 $l'=1$ for $\Phi_{1,3}$ and as a result the first index is unchanged and for the
 second one we use the usual rule of the addition of two spins $l'$ and $1$ which
 leads to $l'$ and $l' \pm 1$, i.e. in terms of $m=2l'+1$ it leads to
  $m$ and $m \pm 2$.  It is also known that only  operators with close
 dimensions  are effectively mixed and then one can conclude \cite{zam1}
 that the 
operator $\Phi_{n,n}$ does not mix with other fields and is the same  in both
 CFT - $M_{k}$ and $M_{k-1}$. This is in perfect agreement with the fact that this
 operator corresponds to the matter field with equal charges with respect to $A$ and $C$
 fields, so after exchange it will be the same operator.   If we consider the pair
 $\Phi_{n,n\pm 1}$ near the UV fixed point $M_{k}$ it will transform along the
 flow into the pair $\Phi_{n\pm 1,n}$. Again this is nothing but an effective
 exchange of the $A$ and $C$ fields. Our last example is a triple of fields
 $\Phi_{n,n \pm 2}$ and $\partial_z\partial_{\bar{z}}\Phi_{n,n}$  which  transforms
   into another triple in the infrared - $\Phi_{n\pm 2,n}$ and 
  $\partial_z\partial_{\bar{z}}\Phi_{n,n}$. One can study in an analogous manner the
 renormalizations of the other fields $\Phi_{n,m}$ - and  all the time we'll see that
 it is nothing but the  exchange $n \leftrightarrow m$, i.e. nothing but
  flip of the sign in the  $SU(2)_{1}$ sector.

   Now let us demonstrate that the renormalization (\ref{renormcscoset}) takes place
 indeed. It is clear that the only matter fields which can contribute to
 the parity violating CS terms are fermions or topologically massive gauge bosons.
 The operator $\Phi_{1,3}$ has a conformal dimension $\Delta_{1,3} = 1
 - 2/(k+3)$ (the same for $\bar{\Delta}$)
 which can be written in the form
\begin{equation} 
 \Delta_{1,3} = -\frac{1}{2} + 1  + \left(\frac{1}{2}  - \frac{2}{k+3}\right)
\label{NewDelta}
\end{equation}
 more convenient for our purposes. We also have to remind 
 that  conformal dimension defines the spin of the matter field.
 So   equation (\ref{NewDelta}) tells us
 that our matter field is a combination of  non-interacting
 fermion (the first factor $-1/2$),  a vector (the second factor $1$) 
 and a fermion in an adjoint representation of $SU(2)_{k+1}$.
 It is known (see \cite{acks} and references therein)
 that a fermion in a representation $R$ induces the Chern-Simons term
 $ {\rm sgn}(m)T(R) S_{CS}$  with the sign depending on the fermion mass $m$ and
$T(R)$ defined as $Tr_{R} T^a T^b = T(R) \delta^{ab}$. For the adjoint representation
 of the $SU(N)$ group $T(G) = N$ and for $SU(2)$ one  gets the shift 
 $k \rightarrow k +  2 {\rm sgn }(m) $. As it has been discussed earlier the $M_{k}$
 model is a UV fixed point where the matter  contribution to the total
 Chern-Simons coefficients must be taken into account. Adjusting the mass $m$
  to be negative  we can get  $-(k+1)$ in front
 of $S_{CS}\{C\}$ in (\ref{cscoset})  as $ - (k-1) -2$, i.e. without matter
 the  bare CS coefficient was $-(k-1)$ - precisely what we need !

 Let us talk about $B$ field corresponding to the $SU(2)_{1}$ factor. We found
 that the matter field interacting with $B$ field must have spin one, i.e., it
 is a vector field $V$ itself and to get a P-odd vertex one can write the 
 lagrangian
\begin{equation}
S_V = \int_{\cal M} \epsilon^{\mu\nu\lambda}Tr V_{\mu}\left(\partial_{\nu}+
B_{\nu}\right) V_{\lambda} = 
\int_{\cal M} \epsilon^{\mu\nu\lambda}Tr V_{\mu}D_{\nu}(B) V_{\lambda}
\label{Vaction}
\end{equation}
 and the induced action after integrating over $V$ will be
 given by $ \ln ~det D(B)$. The same operator appears in the CS action
 (\ref{csaction})  when we split the 
field $A = B + V$, where $B$ is a background
 classical field and $V$ decsribes quantum fluctuations. It is known 
 \cite{Wita} that after proper regularization one can obtain from this
 determinant the correction to the classical CS action leading to the
 shift $k  \rightarrow k + T(G)~{\rm  sgn}(k)$. This shift is obtained
 using  a regularization equivalent to adding a  $Tr F^2$ term to the Chern-Simons
 action which transforms topological Chern-Simons theory into
  topologically massive gauge theory (TMGT) \cite{djt}. This regularization is not
 unique and one can choose another
 regularization and  get a  shift $k  \rightarrow k - T(G) {\rm sgn}( k)$
  (see \cite{shifman} for detailed discussion). Actually the sign is dictated
 by the sign of the mass of the
massive vector boson propagating inside the loop. In TMGT
 this sign is given by the sign of  $k$, but in our case the sign of the  $V$ 
 particles mass is in our hands.  Choosing it opposite to the sign of initial
 $k$  one gets the  shift $k \rightarrow k - 2{\rm sgn}(k)$. Let us  now take 
 $k = -1$ before integration over matter field. 
 After  the  integration over the matter fields we
 get the new $k = -1 + 2 = 1$. This is the sign flip of the $SU(2)_{1}$ factor
 which   was   the most crucial element in  our construction. Let us also
 mention  that  the two  different choices of the mass sign for the $V$ field 
  correspond to either $SU(2) \times SU(2)$ or $SO(1,3)$ symmetry of the
 total action for  the $B$ and $V$ fields. In  this construction $B$ plays
 the role of  the vector field (rotations)  and $V$ plays the role of the
 axial field (boosts).  The choice we  make here corresponds to  a $SO(1,3)$ 
 symmetry.

 As a result we have
demonstrated that the transformation (\ref{renormcscoset})  is induced
 by the three-dimensional matter corresponding to the $\Phi_{1,3}$ field.
  One can repeat the same analysis using $\Phi_{3,1}$ field and demonstrate how
 in this case the   transformation inverse to (\ref{renormcscoset}) takes place.
 The anomalous dimension of the $\Phi_{3,1}$ operator in  the $M_{k-1}$ model  can
  be written by analogy with (\ref{NewDelta}) as
\begin{equation} 
 \Delta_{3,1} =   - \left(\frac{1}{2}  - \frac{2}{k+1}\right)
 + 1 - \frac{1}{2}
\label{NewDelta31}
\end{equation}
Now we have the adjoint fermion with respect to the first $SU(2)_{k-1}$ group
 corresponding to the field $\tilde{A} = C$. It is clear  also that the sign of the
 fermion mass now is opposite (because of  the parity transformation in 
 (\ref{renormcscoset})). So the fermion contribution will be $+2$ and matter
 contribution will transform $k-1$ into $k+1$ - and at the same time we shall get the
 same  sign flip for the
$B$ field, i.e. it describes the transition from $M_{k-1}$
 to $M_{k}$.

 Let us briefly discuss  how  this construction can be generalized to  describe the
 RG flows  between minimal $N=1$ superconformal models \cite{n1} $SM_{k}$.
 These   models have central charge (note that  the central charge
 of SUSY $SU(2)_{k}$ WZNW model is $c_{SSU(2)_k} =
 3/2 + 3(k-2)/k$ and $k \geq 2$) 
\begin{equation}
c_{SM_{k}} = c_{SSU(2)_k}+c_{SSU(2)_2}-c_{SSU(2)_{k+2}} = \frac{3}{2}
\left(1 -\frac{8}{k(k+2)}\right)
\label{sminmodcentrcharge}
\end{equation}
and can be described by the $SU(2)_{k}\times SU(2)_{2}/SU(2)_{k+2}$ 
 coset construction \cite{gko}. The dimensions of the primary superfields are given by
\begin{eqnarray}
\Delta_{2l+1,2l'+1} = \frac{[(k+2)(2l+1) - k(2l'+1)]^2 - 4}{8k(k+2)}
 + \frac{\epsilon}{8} =
 \nonumber \\
\frac{n^2}{2} + \frac{l(l+1)}{k} - \frac{l'(l'+1)}{k+2} +
\frac{\epsilon}{8};~~~~n = l'-l
\label{superDelta}
\end{eqnarray} 
where for  Neveu-Schwartz (NS) superfields $n$ is an integer and $\epsilon =0$,
 and  for  Ramond (R) fields $n$ is a half-integer and $\epsilon = 1/2$. A 
primary
 NS superfield ${\bf \Phi}_{p,q}(\xi, \theta, \bar{\theta}) = \Phi_{p,q}(\xi)
 + \theta \Psi_{p,q}(\xi) + \bar{\theta} \bar{\Psi}_{p,q}(\xi) + 
 \tilde{\Phi}_{p,q}(\xi)$ has two boson comnponents $\Phi_{p,q}$ and
 $ \tilde{\Phi}_{p,q}$
 with dimensions $(\Delta_{p,q}, \bar{\Delta}_{p,q})$ and 
 $(\Delta_{p,q} +1/2, \bar{\Delta}_{p,q}+1/2)$. It is the field $ \tilde{\Phi}_{1,3}$
 which has dimension $\Delta_{1,3} +1/2 = 1 - 2/(k+2)$  near $1$. The  RG flow
 corresponding to the $\tilde{\Phi}_{1,3}$ deformations was studied  in 
 \cite{supermm1}, \cite{supermm2}. It was shown  in  \cite{supermm1} that this
 flow    describes  the transition $SM_{k} \rightarrow SM_{k-2}$, and in \cite{supermm2}
  this result was confirmed  using the Landau-Ginzburg  description of the
 superconformal minimal models. The fact that $\Delta~k = 0({\rm mod}~2)$ is due
 to the fact that the supersymmetry is not broken by this deformation. 
The Witten index  $Tr(-1)^{F}$  then must be the same for both  UV and IR
  fixed points and  because it equals to $(-1)^{k}$ for $SM_{k}$ one  must conclude
 that $k \rightarrow k-2$ is  the minimal possible change of $k$. The renormalization
 of the superfields ${\bf \Phi}_{p,q}$ along the RG trajectory 
$SM_{k} \rightarrow SM_{k-2}$ was found to be of the same type as in the minimal
 model case ${\bf \Phi}_{p,q}^{(k)} \rightarrow {\bf \Phi}_{q,p}^{(k-2)}$.

  The three-dimensional description of the $SM_{k}$ model  is given by the $N=1$ SUSY
 CS theory with  an action
\begin{equation}
kS_{SUSY~CS}\{A\} + 2 S_{SUSY~CS}\{B\} - (k+2)S_{SUSY~CS}\{C\}
\label{susycscoset}
\end{equation} 
 and it is quite clear now that to describe $SM_{k} \rightarrow SM_{k-2}$  transition
 we must do the following: $2 \rightarrow 2-4 = -2$ in the second term and
 $k + 2  \rightarrow k + 2 -4 = k-2$ in the third one. After making  a  parity
 transformation  we shall exchange the roles of $A$ and $B$ fields and finally
 get the three-dimensional description of $SM_{k-2}$. The only difference
  in comparison with the non-SUSY case is that we must subtract $4$ and not $2$. But
 this is due to the fact that  we add a  supersymmetric matter  now. So we have to
 add supermultiplets of fermions and topologically  massive vector bosons - and 
 each of them, as it  has been shown, contributes $2$ to the  renormalization of the
 corresponding CS coefficient, so total contribution will be  $4$ in both cases.
 Thus  we see that our three-dimensional  picture is valid in $N=1$ superconformal
  case also.
 
 In conclusion let us discuss several important questions which have to be considered
 next. First of all it will be interesting to study the $N=2$ case which is important
   in the (super)string theory, especially for  description of transitions between
 different Calabi-Yau manifolds and RG flows on moduli spaces.
  Let us note that  because of the sign flip nature of the three-dimensional transition
  we have at some scale the Chern-Simons theory of the type $[k,0;k]$ which corresponds
 to the coset $(SU(2)_{k}/SU(2)_{k}) \times SU(2)_{0}$. It is interesting to know
 if the appearence of the topological CFT $G/G$ as well as $SU(2)_{0}$ will be a generic 
 feature or not. This may be important for  the  membrane interpretation
  of the conifold transition \cite{strominger}.  Another important
 problem is the three-dimensional analog of the Zamolodchikov $c$-function \cite{zam1}.
 At the fixed point $c$ equals the corresponding central charge and its 
 three-dimensional
 interpretation \cite{Witb}, \cite{membrane}, \cite{kogan2}
  is the gravitational Chern-Simons coefficient in the induced
 topologically massive gravity \cite{djt}. However  it is absolutely unclear
 what is the three-dimensional description of the whole $c$-function and
 this question as well as many others definitely deserve  to be further explored.

 \medskip

\noindent
{\bf Acknowledgments.} 	

\noindent
 I would like to thank J.Cardy for  very  useful  discussions and  suggestions,
 particularly for suggesting  to check the renormalization of the $\Phi_{n,m}$ fields
  in the three-dimensional approach  and R. Szabo for very stimulating and
 important  discussions. 
  I benefited a lot from very interesting discussions
 with L. Cooper, A. Lewis, N. Mavromatos, G. Ross, 
  M. Shifman,  A. Tsvelik,  A. Vainshtein and J. Wheater. This work was
 partly supported by PPARC travel grant and TPI, University of Minnesota.

\vspace{12mm}

\setcounter{section}{1}
\setcounter{equation}{0}
\setcounter{subsection}{0}
\setcounter{footnote}{0}

{\renewcommand{\Large}{\normalsize}
\newcommand{\NPB}[1]{{\sl Nucl.~Phys.}~{\bf B#1}}
\newcommand{\Ann}[1]{{\sl Ann.~Phys.}~{\bf #1}}
\newcommand{\CMP}[1]{{\sl Commun.~Math.~Phys.}~{\bf #1}}
\newcommand{\PLB}[1]{{\sl Phys.~Lett.}~{\bf B#1}}
\newcommand{\PRL}[1]{{\sl Phys.~Rev.~Lett.}~{\bf #1}}
\newcommand{\PTP}[1]{{\sl Prog.~Theor.~Phys.}~{\bf #1}}
\newcommand{\MPLA}[1]{{\sl Mod.~Phys.~Lett.}~{\bf A#1}}
\newcommand{\IJMP}[1]{{\sl Int.~J.~Mod.~Phys.}~{\bf A#1}}
\newcommand{\IJMPB}[1]{{\sl Int.~J.~Mod.~Phys.}~{\bf B#1}}
\newcommand{\CQG}[1]{{\sl Class.~Quant.~Grav.}~{\bf #1}}
\newcommand{\PRD}[1]{{\sl Phys.~Rev.}~{\bf D#1}}
\newcommand{\PRB}[1]{{\sl Phys.~Rev.}~{\bf B#1}}
\newcommand{\JMP}[1]{{\sl J.~Math.~Phys.}~{\bf #1}}

\end{document}